\documentclass[
  aps,pra,
  superscriptaddress,
  longbibliography,
  12pt
  ]{revtex4-2}

\usepackage{graphicx}
\usepackage{braket}
\usepackage{hyperref}
\usepackage{physics}   
\usepackage{amsmath,amsthm,amssymb}
\usepackage{mathtools}
\usepackage{bm}
\usepackage[normalem]{ulem}

\usepackage{calc}
\usepackage[greek,english]{babel}
\usepackage{tikz}

\newtheorem{theorem}{Theorem}
\newtheorem*{theorem*}{Theorem}
\newtheorem{proposition}[theorem]{Proposition}
\newtheorem{lemma}[theorem]{Lemma}
\newtheorem*{lemma*}{Lemma}
\newtheorem{corollary}[theorem]{Corollary}

\newif\ifcmnt
\cmnttrue 
\ifdefined\cmntsoff\cmntfalse\fi

\ifcmnt
    \providecommand{\aucmnt}[1]{#1}

\else
    \providecommand{\aucmnt}[1]{}

\fi

\providecommand{\ignore}[1]{}

\newcommand\bluesout{\bgroup\markoverwith{\textcolor{Cerulean}{\rule[0.5ex]{2pt}{1pt}}}\ULon}

\begin{document}

\title{Reversing quantum dynamics with near-optimal quantum and classical fidelity}

\author{Howard Barnum}
\affiliation{Los Alamos National Laboratory, Los Alamos, NM 87545, USA}
\affiliation{Dept. of Computer Science, University of Bristol, Bristol BS8 1UB, UK}

\author{Emanuel Knill}
\affiliation{Los Alamos National Laboratory, Los Alamos, NM 87545, USA}

\begin{abstract}
  We consider the problem of reversing quantum dynamics, with the goal
  of preserving an initial state's quantum entanglement or classical
  correlation with a reference system. We exhibit an approximate
  reversal operation, adapted to the initial density operator and the
  ``noise'' dynamics to be reversed. We show that its error in
  preserving either quantum or classical information is no more than
  twice that of the optimal reversal operation.  Applications to
  quantum algorithms and information transmission are discussed.
\end{abstract}
\maketitle

\section{Introduction}

Counteracting the effects on quantum systems of noise generated by
interaction with an environment is a central problem for the emerging
field of quantum information processing. Its solution can be expected
to have applications in quantum computation, precision measurement,
and information transmission. In this paper we exhibit a reversal
operation which takes account of both the noise and the initial
density operator, to achieve near-optimal preservation of the initial
density operator's quantum entanglement or classical correlation with
a reference system. We work throughout with finite-dimensional systems
although we expect that the results generalize to infinite-dimensional
ones, at least those with separable Hilbert spaces.

Section~\ref{sec2} reviews some of the theory of completely positive
(CP) linear maps on spaces of operators on finite-dimensional Hilbert
spaces. When these are trace-nonincreasing, they are also known in the
quantum information/computation community as ``quantum operations.''
The only notation not standard in quantum information theory is our
writing \(\mathcal{A}\sim\{A_{i}\}_{i}\) to indicate that a CP-map acts as
\(\mathcal{A}(\rho)= \sum_{i} A_{i}\rho A_{i}^{\dagger}\) . Section~\ref{sec3}
reviews measures of fidelity commonly used in quantum information
theory to quantify the effect of noise and information-processing
operations on a system's state, or an ensemble of system states. In
addition, it motivates the particular measures we use, and indicates
their connection to fidelity measures used in classical
information-transmission theory. The measures we use concentrate on
preserving the correlation or entanglement of a mixed state with a
reference system. We explain why they are nevertheless relevant to the
preservation of ensembles of pure states. This section includes
several formal definitions required in the rest of the
paper. Sections~\ref{sec4} and~\ref{sec6} are the core of the paper,
containing the two main results. In Sec.~\ref{sec4} the near-optimal
reversal operation is defined, and we prove one of our main results:
that the reversal operation's error is no worse than twice that of the
optimal reversal operation, for either quantum or classical
information. Section~\ref{sec5} discusses the relationship of the
reversal operation to a known near-optimal method of recovering
classical information encoded in an ensemble of density operators, the
``pretty good measurement''. When the noise can be viewed as encoding
classical information in a completely decohered ensemble of density
operators, our reversal operation can be viewed as performing the
pretty-good measurement to distinguish these density operators. But
when the noise operation may be viewed as encoding classical
information into these density operators in a way which leaves some
coherence between them, our reversal operation writes a nearly optimal
estimate of the density operator into a set of orthogonal states
without decohering these states as a measurement
would. Section~\ref{sec6} proves the other main result of the paper: a
useful lower bound on the fidelity with which the reversal operation
recovers classical information. This bound can be more tractable than
the reversal operation's fidelity itself, and has proven useful, for
example, in discovering upper bounds on quantum query complexity. This
and other applications are touched on in the concluding section.

\section{Quantum noise operations}
\label{sec2}

We model quantum noise in a finite dimensional system \(Q\) by the
most general dynamics that can arise via a unitary interaction
\(U^{QE}\) with an environment \(E\) initially independent of the
system, in the sense that the initial joint density matrix of system
and environment is a product \(\rho^{Q}\otimes \sigma^{E}\). The state
of the system after the dynamics is obtained by tracing out the
environment after \(U^{QE}\) has been applied.  We also model
operations performed on a system in an attempt to counteract noise by
such dynamics coupled to different, initially independent environment.
For our purposes, where we ultimately care about the effect of
operations on the system and perhaps on its entanglement with systems
other than the environment, we may model operations with the
environments starting in a pure state. This is because the effect on
the system of a unitary interaction \(U^{QE}\) with a mixed state
\(\sigma\) of \(E\) can be replicated by an interaction
\(U^{QE_{1}} \otimes I^{E_{2}}\) with a pure state \(\ket{0}\) of an
environment \(E=E_{1}\otimes E_{2}\) , whose partial trace onto
\(E_{1}\) is equal to \(\sigma\).

The most general dynamics thus obtained are trace-preserving
completely positive maps \(\mathcal{A}\) and can be represented in the
form \(\mathcal{A}:\rho\mapsto\sum_{i}A_{i}\rho A_{i}^{\dagger}\) for
\(\rho\) in the space of linear operators on \(Q\), where
\(\{A_{i}\}_{i}\) is a finite family of linear operators on \(Q\). We
normally leave the number of operators in the family unspecified.  The
condition that the dynamics is trace preserving implies that
\(\sum_{i}{A_{i}}^{\dagger}A_{i}=I\). If the dynamics arises from
tracing out the environment from the state
\(U^{QE}\qty(\rho\otimes\ketbra{0^{E}}){U^{QE}}^{\dagger}\), then
\begin{align}
  A_{i}= \bra{i^{E}}U^{QE}\ket{0^{E}}
  \label{eq1}
\end{align}
are the ``operator matrix elements'' of the unitary interaction, with
the elements computed between the environment initial state and states
\(\ket{i^E}\) forming an orthonormal basis for the environment.  In
general we define the operator matrix element
\(\bra{\xi^{B}}O^{AB}\ket{\phi}^{B}\) of an operator \(O^{AB}\) as
follows: Let the matrix elements of \(O\) in some tensor product basis
be \(O_{ij,i'j'}\) and let \(\xi_{j}\) and \(\phi_{j}\) be the
components of \(\ket{\xi}\) and \(\ket{\phi}\) in the basis
\(\ket{j^{B}}\). Then the operator matrix element
\(\bra{\xi^{B}}O^{AB}\ket{\phi}^{B}\) is the operator \(X\) on \(A\)
whose matrix elements in the basis \(\ket{i^{A}}\) are given by
\begin{align}
  \bra{i}X\ket{i'} = \sum_{j,j'}\xi^{*}_{j}\phi_{j'}O_{ij,i'j'}.
\end{align}

States \(A_{i}\ket{\psi}\) obtained by applying \(A_{i}\) to an
initial state \(\ket{\psi}\) of \(Q\) are often called the
(unnormalized) ``relative state'' of the system \(Q\), in the sense
that it is relative to the environment basis state \(\ket{i^{E}}\)
after the unitary interaction. The overall evolution of system and
environment is
\begin{align}
  \ket{\psi^{Q}}\ket{0^{E}}\rightarrow U^{QE}\ket{\psi^{Q}}\ket{0^{E}} =
  \sum_{i}A_{i}^{Q}\ket{\psi^{Q}}\ket{i^{E}}
\end{align}
That the state on the right-hand side is normalized for normalized
input states is equivalent to the requirement that
\(\sum_{i}{A_{i}}^{\dagger}A_{i}=I\) for trace-preserving quantum
operations.  Quantum operations in general are trace-nonincreasing,
which is equivalent to the requirement that
\(\sum_{i}{A_{i}}^{\dagger}A_{i}\leq I\).  Quantum operations
\(\mathcal{A}\) satisfy that if \(F\leq G\) are positive semidefinite
operators, than \(\mathcal{A}(F)\leq\mathcal{A}(G)\).

If the environment is considered as a measuring apparatus with the
states of the basis \(\ket{i^{E}}\) its ``pointer variables''~\cite{Zurek1981}
corresponding to different measurement results, then
\(A_{i}\ket{\psi}\) is the state of the system conditional on the
apparatus measurement outcome \(i\) after coupling. The squared norm
of \(A_{i}\ket{\psi}\) is the probability of the measurement outcome
\(i\). The overall operation \(\mathcal{A}\) represents the dynamics
of the measurement averaged over measurement results, corresponding to
the sitatuation where the measurement results are ignored. Of course,
a trace-preserving operation need not arise from an actual ignored
measurement in this manner, but it may always be viewed this way, in
terms of averaging over a notional readout of the environment in some
``pointer basis'', if desired.

The \(\mathcal{A}_{i}\) are said to form a decomposition of
\(\mathcal{A}\), for which we use the notation
\(\mathcal{A}\sim \{A_{i}\}_{i}\).  Using a different orthonormal ``pointer
basis''~\cite{Zurek1981} for the environment in Eq.~\eqref{eq1}
results in a different decomposition of the same operation. Two such
pointer bases are related by a unitary transformation of the
environment. The corresponding decompositions of the operation are
related by the unitary \(V\)
that takes the pointer basis used to define the \(A_{j}\) to that used to define the
\(B_{i}\):
\begin{align}
  B_{i}=\sum_{j}{V_{ji}}^{*}A_{j},
\end{align}
where \(V_{ji}=\bra{j^{E}}V\ket{i^{E}}\) are the entries of \(V\) in
the pointer basis \(\ket{i^{E}}\) used to define the \(A_{j}\).

\section{Quantifying the effects of noise}
\label{sec3}

Let us consider how to quantify the effects of a noise operation on
information contained in a quantum system, using the classical Shannon
theory of information transmission as a guide. The information we have
about a set of classical alternatives indexed by \(i\) of a local
system \(T\) is usually measured by the entropy of the probabilities
we ascribe to the alternatives:
\(H(\bm{p})\coloneqq -\sum_{i}p_{i}\log(p_{i})\). Suppose we want to
be able to transmit or store the classical states \(i\), which will be
presented to us with the probabilities \(p_{i}\) . We encode them on a
system \(T\), while keeping a reference copy with us, on a system
\(R\). In transmission, \(T\) is then affected by noise. We might
define perfect success as the preservation of the initial perfect
correlation between \(R\) and \(T\). In terms of the probability
measures, we want the initial and final joint mixed states of
reference \(R\) and system \(T\) both to be described by
\(p(i^{R}, j^{T}) = p_{i}\delta_{i^{R}j^{T}}\).  If our actual,
noise-affected transmission results in the mixed state with joint
probability distribution \(q(i^{R},j^{T})\), we could measure how well
we have succeeded by looking at the distance between
\(p(i^{R},j^{T})\) and \(q(i^{R}, j^{T})\) using some standard measure
of distance between probability distributions. To give examples of
such distances, let
\(\bm{r}=(r_{1},\ldots, r_{n}), \bm{s}=(s_{1},\ldots,s_{n})\) be
probabilities on a common set of classical alternatives labeled
\(1, \ldots ,n\). We can, for example, measure the distance between
\(r\) and \(s\) by the classical infidelity
\(B(\bm{r},\bm{s})\coloneqq
1-\qty(\sum_{k}r_{k}^{1/2}s_{k}^{1/2})^{2}\), which is a function of
the Bhattacharyya coefficient.  The total variation distance
\(\frac{1}{2}\sum_{ij}|r_{k}-s_{k}|\) is another reasonable choice. We
assume that \(q(i,j)=p_{i}t(j|i)\), where \(t(j|i)\) is a stochastic
matrix of transition probabilities describing the noise. Using the
classical infidelity to measure how well the original correlation has
been preserved in the above-given example gives
\begin{align}
  1-\qty(\sum_{i,j}\delta_{ij}p_{i}^{1/2}q(i,j)^{1/2})^{2}
  = 1-\qty(\sum_{i}p_{i}^{1/2}q(i,i)^{1/2})^{2} = 1-\qty(\sum_{i}p_{i}t(i|i)^{1/2})^{2}.
\end{align}
Using the total variation distance gives
\begin{align}
  \frac{1}{2}\sum_{ij}|\delta_{ij}p_{i}-p_{i}t(j|i)|
  &= \sum_{ij: p_{i}t(j|i)> p_{i}}p_{i}t(j|i) -\delta_{ij}p_{i}
    \notag\\
  &=\sum_{ij: j\ne i}p_{i}t(j|i)
    \notag\\
  &= \sum_{i}p_{i}\sum_{j\ne i}t(j|i).
\end{align}
The expression \(\sum_{j\ne i}t(j|i)\) is the probability that, if
\(i\) is sent, a different message is received.  The total variation
distance therefore is the same as the familiar ``error probability''
criterion for channel transmission: the average, over messages \(i\),
of the probability that \(i\) is sent but a different message is
received. This can be interpreted as an average pure-state infidelity
where the reference pure state for input \(i\) is \(\ket{i}\), but we
have shown that it can be rewritten as a distance from a mixed state
representing perfect initial correlation.  For information theoretic
purposes, the classical infidelity and the total variation distance
have closely related behaviors.  The relationships can be determined
from the inequalities given in Ref.~\cite{fuchs}, Prop. 5.

We have gone through this analysis to indicate that there is a
classical analog of the approach we here take to quantum as well as
classical fidelity: investigating the effects of noise on a channel,
which we may want to use to transmit pure states, by looking at its
effect on a state whose marginal distribution on the noise-affected
system is a mixed state corresponding to the ensemble of possible
pure-state messages. Thus an approach which might be viewed as
unnecessarily concerned with the fidelity of mixed states, provides a
good fidelity criterion for the problem of transmitting pure states
which are supplied to the channel with probabilities \(p_{i}\), where
the fidelity criterion is in terms of the idea of preservation of
correlation with a possibly notional reference system.

We now apply the classical analysis given above to the problem of
transmitting quantum information, specifically the problem of
evaluating the fidelity with which a given operation preserves quantum
information. We resume consideration of the noise operation
\(\mathcal{A}\), which is the analog of the classical stochastic
channel matrix \(t(i|j)\) presented earlier. For the purpose of
defining entanglement fidelities, we weaken the trace-preserving
property and consider general quantum operations, namely
trace-nonincreasing, completely positive maps \(\mathcal{A}\).  Now,
rather than preservation of the noise-affected system's correlation
with a reference system, we require preservation of its entanglement.
Just as the classical reference system could be completely notional,
the quantum reference system may also be taken to be completely
notional. In both cases, the reference system is introduced because
high-fidelity preservation of correlation with the reference system
implies high-fidelity preservation of states on the noise-affected
system when they are supplied to the channel with the marginal
distribution \(p_{i}\). When \(\mathcal{A}\) acts on \(Q\), a
state
\(\ket{\psi_{0}^{RQ}}\coloneqq
\sum_{i}\sqrt{p_{i}}\ket{i^{R}}\ket{i^{Q}}\) entangled with a reference system \(R\)
evolves as
\begin{align}
  \ket{\Psi_{0}}\coloneqq\sum_{i}\sqrt{p_{i}}\ket{i^{R}}\ket{i^{Q}}\ket{0^{E}}
  \rightarrow
  \ket{\Psi_{f}}\coloneqq\sum_{ij}\sqrt{p_{i}}\ket{i^{R}}A_{j}\ket{i^{Q}}\ket{j^{E}}.
  \label{eq6}
\end{align}
The entanglement fidelity \(F_{e}(\rho,\mathcal{A})\) is defined as
\(\|P_{0}\ket{\Psi_{f}}\|^{2}\), where
\(P_{0}\coloneqq\ket{\psi_{0}^{RQ}}\bra{\psi_{0}^{RQ}}\otimes
I^{E}\)~\cite{Schumacher1996}.  Thus \(F_{e}\) is the squared norm of
the projection of the final state in Eq.~\eqref{eq6} onto the subspace
associated with the initial entangled state
\(\ket{\psi_{0}^{RQ}}\). It depends only on
\(\rho\coloneqq\sum_{i}p_{i}\ket{i^{Q}}\bra{i^{Q}}\), and satisfies
\begin{align}
  F_{e}(\rho,\mathcal{A}) = \sum_{i}|\tr A_{i}\rho|^{2}.
  \label{eqfe}
\end{align}
We can also define the input-output fidelity for a pure state
\(\ketbra{\psi}\) as
\(F(\ket{\psi},\mathcal{A}) =
\bra{\psi}\mathcal{A}(\ketbra{\psi})\ket{\psi}\), which is the
fidelity of the final state of \(Q\) onto its initial state.  As in
the classical case with correlation, using entanglement for the
quantum case does not reflect a fixation on entanglement, but also
provides an appropriate fidelity criterion even when it is
preservation of pure states on the system \(Q\) that we are concerned
with, and no entangled system \(R\) in fact exists. This is because
the entanglement fidelity \(F_{e}(\rho^{Q},\mathcal{A})\) is a lower
bound on the average input–output fidelity
\(\sum_{i}q_{i}F(\ket{\psi_{i}},\mathcal{A})\) for any pure-state
ensemble on \(Q\) consisting of \(\ket{\psi_{i}}\) with probability
\(q_{i}\) for which its density operator
\(\sigma^{Q}=\sum_{i}q_{i}\ketbra{\psi_{i}}\) satisfies
\(\sigma^{Q}=\rho^{Q}\).  While this is a good lower bound on the
input–output fidelities of pure-state ensembles, it is not in general
a tight lower bound on the input–output fidelity. Thus if one is
really interested only in the effect of a quantum channel on a
particular ensemble, one might get fidelities much higher than the
entanglement fidelity suggests. We will be interested in one such
case, the input–output fidelity for an orthonormal set of states,
related to the ability to preserve essentially classical information
coded in a quantum channel. In order to treat this case by utilizing
entanglement fidelities, we introduce a notion which generalizes both:
average entanglement fidelity.

For an ensemble \(E=\{p_{i},\rho_{i}\}_{i}\) where state \(\rho_{i}\)
occurs with probability \(p_i\), we define the average entanglement
fidelity by
\begin{align}
\bar F_{e}(E,\mathcal{A}) \coloneqq
  \sum_{i}p_{i}F_{e}(\rho_{i},\mathcal{A}).
  \label{eqafe} 
\end{align}
A special case is
\begin{align}
  F_{\text{cl}}(\rho,\mathcal{A})
  \coloneqq \sum_{i}p_{i}\bra{i}\mathcal{A}(\ketbra{i})\ket{i}
  = \bar F_{e}(\{p_{i},\ketbra{i}\}_{i},\mathcal{A}),
\end{align}
where the \(\ket{i}\) form an eigenbasis of \(\rho\) with
\(\rho=\sum_{i}p_{i}\ketbra{i}\).  This is the classical fidelity for
the classical information of the ensemble of orthogonal eigenstates of
the input density operator \(\rho\). Another special case is an
ensemble consisting of a single density operator \(\rho\). In this
case \(\bar F_{e}\) is just \(F_{e}(\rho,\mathcal{A})\).

The average entanglement fidelity can equivalently be expressed as the
norm squared of the projection of the overall final state onto the
subspace in which entangled states \(\ket{\psi_{i}^{RQ}}\)
representing the initial ensemble are correctly correlated with
orthogonal states of an additional reference system \(S\). There are
different choices of states of \(RQS\) that have such correct
correlations. One is
\(\ket{\psi_{0}^{RQS}}
=\sum_{i}\sqrt{p_{i}}\ket{\psi_{i}^{RQ}}\ket{i^{S}}\), which
represents the  correlation by entanglement with
\(S\). Another is the mixed state
\(\ketbra{\psi_{0}^{RQS}}=\sum_{i}p_{i}\ketbra{\psi_{i}^{RQ}}\otimes
\ketbra{i^{S}}\), for which the correlation is classical.  With
either choice, an equivalent expression for average entanglement
fidelity is
\begin{align}
  \bar F_{e} &= \|P_{c}\otimes I^{E}\ket{\Psi_{f}}\|^{2}
               \notag\\
             &=\tr P_{c}\mathcal{I}^{RS}\otimes\mathcal{A}\qty(\ketbra{\psi_{0}^{RQS}}),
\end{align}
where
\(P_{c}\coloneqq
\sum_{i}\ketbra{i^{s}}\otimes\ketbra{\psi_{i}^{RQ}}\).  This
expression does not depend on whether the correlation with \(S\) in
\(\ketbra{\psi_{0}^{RQS}}\) is obtained by entanglement or merely
classically.

For example, consider the special case of \(F_{\text{cl}}(\rho,\mathcal{A})\) . Here the
reference system \(R\) plays no role as the \(\rho_{i}\) are
pure. After suppressing \(R\),
\(P_{c}=\sum_{i}\ket{i^{S}}\ket{i^{Q}}\bra{i^{Q}}\bra{i^{S}}\).  \(S\)
contains a record of the classical information sent. \(S\) and \(Q\)
may be supposed to be either entangled or classically correlated, with
\(\rho_{0}^{SQ}=\sum_{i,j}\sqrt{p_{i}p_{j}}\ketbra{i^{S}}{j^{S}}\otimes\ketbra{i^{Q}}{j^{Q}}\)
or \(\rho_{0}^{SQ}=\sum_{i}p_{i}\ketbra{i^{S}}\otimes\ketbra{i^{Q}}\),
for orthonormal system and reference bases \(\ket{i}\), where the
system basis is the eigenbasis of \(\rho\). In either case, computing
the probability \(\tr(P_{c}\rho_{f}^{SQ})\) that the final
system-reference state falls into the subspace in which system and
reference exhibit perfect classical correlation in the desired bases,
gives the classical fidelity \(F_{\text{cl}}(\rho,\mathcal{A})\).

\section{The reversal operation}
\label{sec4}

We motivate the definition of the near-optimal reversal operation
\(\mathcal{R}_{\mathcal{A},\rho}\) by considering operations \(\mathcal{A}\) that
are perfectly reversible on a ``code'' subspace \(C\). Let \(P_{C}\)
be the projector onto \(C\). Perfectly reversible operations have a
decomposition \(A_{i}\) for which \(A_{i}P_{C}=\sqrt{p_{i}}W_{i}\) for
some probabilities \(p_{i}\), where the \(W_{i}\) are isometries from
\(C\) into orthogonal subspaces, which means that
\(W_{i}^{\dagger}W_{j}=\delta_{ij}P_{C}\)~\cite{KnillLaflamme1997,Nielsen1998}.
Intuitively, this means that as far as its action on the code subspace
is concerned, the operation just maps the state isometrically into one
of a set of mutually orthogonal subspaces. Without loss of generality,
assume that the ranges of the \(W_{i}\) together span the state
space. The reversal operation has a decomposition consisting of the
operators \(W_{i}^{\dagger}=P_{C}A_{i}^{\dagger}/\sqrt{p_{i}}\).
Intuitively, it may be thought of as measurement of which of the
subspaces the state was mapped isometrically to, followed by the
inverse of that isometry to put it correctly back into the code space.
This resembles the adjoint of the restriction \(\mathcal{A}_{C}\) of
\(\mathcal{A}\) to \(C\) with respect to the Hilbert-Schmidt inner
product \((A,B)=\tr A^{\dagger}B\), which is given by
\(\mathcal{A}_{C}^{\dagger}\sim\{P_{C}A_{i}^{\dagger}\}_{i}\).  To get
the reversal operation, the \(\sqrt{p_{i}}\) need to be cancelled,
which also makes the operation trace preserving. The general
definition of the reversal operation
\(\mathcal{R}_{\mathcal{A},\rho}\) is also based on the adjoint,
suitably corrected to ensure that it is trace preserving. The reversal
operation is defined as
\begin{align}
  \mathcal{R}_{\mathcal{A},\rho}
  \sim\{\rho^{1/2}A_{i}^{\dagger}\mathcal{A}(\rho)^{-1/2}\}_{i}\;,
  \label{eq9}
\end{align}
where we assume that \(\mathcal{A}(\rho)\) has full support. In our
analyses, the orthogonal complement of the support of
\(\mathcal{A}(\rho)\) contributes nothing, so there is no loss of
generality in making this assumption.  If we apply this definition to
an operation \(\mathcal{A}\) that is perfectly reversible on a code,
using as our input \(\rho\) any state with support the code, it can be
verified that this definition matches the reversal operation for codes
described previously.

The notation \(\mathcal{R}_{\mathcal{A},\rho}\) does not refer to a specific
decomposition of \(\mathcal{A}\), which is justified by:

\begin{lemma}\label{lem1}
  The definition of the reversal operation
  \(\mathcal{R}(\mathcal{A},\rho)\) is independent of the
  decomposition \(\{A_{i}\}_{i}\) of \(\mathcal{A}\).
\end{lemma}

\begin{proof}
  Proof: Let \(\mathcal{A}\sim\{B_{i}\}_{i}\) be another
  decomposition. By adding null operators to one of the two
  decompositions, we can ensure that both have the same number of
  operators. Adding null operators to \(\{A_{i}\}_{i}\) does not
  change the action of \(\mathcal{R}_{\mathcal{A},\rho}\).  Then there
  exist \(U_{ij}\) such that \(B_{i}=\sum_{j}U_{ij}A_{j}\), where the
  matrix \(U\) with entries \(U_{ij}\) is unitary. The decomposition
  of \(\mathcal{R}_{\mathcal{A},\rho}\) in terms of the
  \(A_{i}^{\dagger}\) given in Eq.~\eqref{eq9} transforms via the
  coefficients of the entry-wise complex conjugate \(U^{*}\) of \(U\)
  into a decomposition given in terms of the \(B_{i}^{\dagger}\). As
  \(U^{*}\) is also unitary, the result is another decomposition of
  the same operation.
\end{proof}

A simple but important property of \(\mathcal{R}_{\mathcal{A},\rho}\) can be verified
directly from the definition:

\begin{proposition}
  \(\mathcal{R}_{\mathcal{A},\rho}(\mathcal{A}(\rho))=\rho\).
\end{proposition}

The operation \(\mathcal{R}_{\mathcal{A},\rho}\) is near-optimal in
the sense given by the following theorem.

\begin{theorem}\label{thm2}
  Let \(E=\{p_{i},\rho_{i}\}_{i}\) be an ensemble of commuting density matrices. Let
  \(\rho=\sum_{i}p_{i}\rho_{i}\).
  Then for every trace-preserving, completely positive map \(\mathcal{R}\),
  \begin{align}
    \bar F_{e}(E, \mathcal{R}_{\mathcal{A},\rho}\mathcal{A}) \geq
    \bar F_{e}(E,\mathcal{R}\mathcal{A})^{2}.
    \label{thm3eq}
  \end{align}
\end{theorem}

As a corollary, if \(\bar F_{e}(E,\mathcal{R}\mathcal{A}) = 1-\eta\),
then
\(\bar F_{e}(E,\mathcal{R}_{\mathcal{A},\rho}\mathcal{A})\geq
(1-\eta)^{2}\geq 1-2\eta\).  That is, the error of
\(\mathcal{R}_{\mathcal{A},\rho}\), defined as one minus the
entanglement fidelity, is never greater than twice that of the best
reversal operation.

\begin{proof}
  Let \(\mathcal{R}\sim \{R_{i}\}_{i}\) be a trace-preserving,
  completely positive map and \(P_{C}\) the projector onto the support
  of \(\rho\).  The average entanglement fidelity of \(\mathcal{R}\)
  is preserved if we replace the \(R_{i}\) by
  \(P_{C}R_{i}\). This results in a quantum operation \(\mathcal{R}\),
  which is trace-nonincreasing.
  With this replacement, there exist
  operators \(B_{i}\) such that
  \begin{align}
    R_{i} = \rho^{1/2}{B_{i}}^{\dagger}\mathcal{A}(\rho)^{-1/2},
    \label{eqrb}
  \end{align}
  namely those defined by
  \({B_{i}}^{\dagger}=\rho^{-1/2}R_{i}\mathcal{A}(\rho)^{1/2}\).
  Inverses such as \(\rho^{-1/2}\) are generalized inverses obtained by
  inverting the non-zero eigenvalues in an eigenbasis while keeping the
  zero eigenvalues.  Let \(\mathcal{B}\sim\{B_{i}\}_{i}\), which need
  not be trace-preserving or trace-nonincreasing.
  However, we have \(\mathcal{B}(\rho)\leq \mathcal{A}(\rho)\):
  Since \(\sum_{i}{R_{i}}^{\dagger}R_{i}\leq I\),
  \begin{align}
    \mathcal{B}(\rho)
    &= \mathcal{A}(\rho)^{1/2}\qty(\sum_i {R_i}^{\dagger}R_{i})\mathcal{A}(\rho)^{1/2}
      \leq \mathcal{A}(\rho),
      \label{eqbarho}
  \end{align}
  where we used the fact that for operators \(F\) and \(G\leq H\),
  \(FGF^{\dagger}\leq FHF^{\dagger}\).  According to
  Eq.~\ref{eqfe} and the definition of average entanglement fidelity,
  \begin{align}
    \bar F_{e}(E,\mathcal{R}\mathcal{A})
    = \sum_{l} p_{l}\sum_{ij}
    |\tr \rho^{1/2}B_{i}^{\dagger}\mathcal{A}(\rho)^{-1/2}A_{j}\rho_{l}|^{2}.
    \label{eq:fera}
  \end{align}
  Define the matrices \(X^{l}\) by
  \(X^{l}_{\,ij}=\tr
  \rho^{1/2}B_{i}^{\dagger}\mathcal{A}(\rho)^{-1/2}A_{j}\rho_{l}\).
  We may assume that the \(X^{l}\) are square, by filling in
  \(\{B_{i}\}_{i}\) or \(\{A\}_{i}\) with null operators, if
  necessary.  To reduce the inner sum to one having just one index, we
  make an \(l\)-dependent choice of operator decompositions
  \(\mathcal{B}\sim\{B^{l}_{\,i}\}_{i}\) and
  \(\mathcal{A}\sim\{A^{l}_{\,i}\}_{i}\) by means of the singular
  value decompositions of the matrices \(X^{l}\) (see for example,
  Ref.~\cite{HornJohnson1985}, Sec. 7.3). Let \(V^{l}\) and \(W^{l}\)
  be unitary transformations such that \({V^{l}}^{\dagger}X^{l}W^{l}\)
  is diagonal.  We define \(B^{l}_{\,i}=\sum_{k}B_{k}V^{l}_{\,ki}\)
  and \(A^{l}_{\,j}=\sum_{k'}A_{k'}W^{l}_{\,k'j}\), so that
  \begin{align}
    \qty({V^{l}}^{\dagger}X^{l}W^{l})_{ij}
    &= \delta_{ij}\sum_{kk'}{V^{l}_{\,ki}}^{*}W^{l}_{\,k'j}
      \tr \rho^{1/2}{B_{k}}^{\dagger}\mathcal{A}(\rho)^{-1/2}A_{k'}\rho_{l}
      \notag\\
    &= \delta_{ij}\tr \rho^{1/2}{B^{l}_{\,i}}^{\dagger}
      \mathcal{A}(\rho)^{-1/2}A^{l}_{\,j}\rho_{l}.
  \end{align}
  The inner sum in Eq.~\eqref{eq:fera} is
  \begin{align}
    \sum_{ij}|X^{l}_{\,ij}|^{2}
    &= \tr\,{X^{l}}^{\dagger} X^{l}
      \notag\\
    &= \tr \qty({V^{l}}^{\dagger} X^{l} W^{l})^{\dagger}\qty({V^{l}}^{\dagger} X^{l} W^{l})
      \notag\\
    &= \sum_{ij}\qty|\qty({V^{l}}^{\dagger}X^{l}W^{l})_{ij}|^{2}
      \notag\\
    &= \sum_{i}|\tr \rho^{1/2}{B^{l}_{\,i}}^{\dagger}\mathcal{A}(\rho)^{-1/2} A^{l}_{\,i}\rho_{l}|^{2}.
  \end{align}
  We also define \(l\)-dependent operator decompositions for
  \(\mathcal{R}\) according to
  \(R^{l}_{\,i}=\sum_{k}R_{k}{V^{l}_{\,ki}}^{*}\), so that
  \({B^{l}_{\,i}}^{\dagger}=\rho^{-1/2}R^{l}_{\,i}\mathcal{A}(\rho)^{1/2}\).

  Define
  \(Y_{li}=p_{l}^{1/4}\mathcal{A}(\rho)^{-1/4}B^{l}_{\,i}\rho^{1/4}\rho_{l}^{1/2}\)
  and
  \(Z_{li}=p_{l}^{1/4}\mathcal{A}(\rho)^{-1/4}A^{l}_{\,i}\rho^{1/4}\rho_{l}^{1/2}\).
  We then obtain the following inequalities, where the steps are explained below:
  \begin{align}
    \bar F_{e}(E,\mathcal{R}\mathcal{A})
    &= \sum_{l}p_{l}\sum_{i}|\tr \rho^{1/2}{B^{l}_{\,i}}^{\dagger}A(\rho)^{-1/2}A^{l}_{\,i}\rho_{l}|^{2}
      \label{s1}\\
    &=\sum_{il}|\tr\, {Y_{li}}^{\dagger}Z_{li}|^{2}
      \label{s2}\\
      &\leq \sum_{il}\tr\, {Z_{li}}^{\dagger}Z_{li}\tr\,{Y_{li}}^{\dagger}Y_{li}
      \label{s3}\\
    &\leq \qty(\sum_{il}|\tr\, {Z_{li}}^{\dagger}Z_{li}|^{2}
      \sum_{i'l'}|\tr\, {Y_{l'i'}}^{\dagger}Y_{l'i'}|^{2})^{1/2}
      \label{s4}\\
    &\leq \qty(\sum_{il}|\tr\, {Z_{li}}^{\dagger}Z_{li}|^{2})^{1/2}
      \label{s5}\\
    &\leq \qty(\sum_{ijl}|\tr\, {Z_{li}}^{\dagger}Z_{lj}|^{2})^{1/2}
      \label{s6}\\
    &=\qty(\sum_{l} p_{l}\sum_{ij}
      |\tr \rho^{1/2}{A^{l}_{\,i}}^{\dagger}\mathcal{A}(\rho)^{-1/2}
      A^{l}_{\,j}\rho_{l}|^{2})^{1/2}
      \label{s7}\\
    &=\bar F_{e}(E,\mathcal{R}_{\mathcal{A},\rho}\mathcal{A})^{1/2}.
      \label{s8}
  \end{align}
  Line~\eqref{s2} uses cyclicity of the trace and, by assumption,
  \([\rho,\rho_{l}]=0\).  Line~\eqref{s3} is the operator Schwarz
  inequality, and line~\eqref{s4} is the vector Schwarz
  inequality. Line~\eqref{s5} follows from
  \(\sum_{i'l'}|\tr {Y_{l'i'}}^{\dagger}Y_{l'i'}|^{2}\leq 1\), which is shown
  below.  Line~\eqref{s6} just adds positive terms under the square
  root.  The remaining equalities involve applying cyclicity of the
  trace and commutativity again, and applying the definitions.

  It remains to establish the inequality required for Line~\eqref{s5}
  above, which follows from
  \begin{align}
    \sum_{i'l'}|\tr\, {Y_{l'i'}}^{\dagger}Y_{l'i'}|^{2}
    &=\sum_{l}p_{l}\sum_{i}
    |\tr\,\rho_{l}\rho^{1/2}{B^{l}_{\,i}}^{\dagger}
      \mathcal{A}(\rho)^{-1/2} B^{l}_{\,i}|^{2}
      \label{t1}\\
    &=\sum_{l} p_{l} \sum_{i}|\tr\,\rho_{l}R^{l}_{\,i}B^{l}_{\,i}|^{2}
      \label{t2}\\
    &=
      \sum_{l}p_{l}\sum_{i}|\tr R^{l}_{\,i}B^{l}_{\,i} \rho_{l}^{1/2}\rho_{l}^{1/2}|^{2}
      \label{t3}\\
    &\leq
    \sum_{l}p_{l}\sum_{i}\tr R^{l}_{\,i}B^{l}_{\,i} \rho_{l} {B^{l}_{\,i}}^{\dagger}
    {R^{l}_{\,i}}^{\dagger} \tr \rho_{l}
      \label{t4}\\
    &=
    \sum_{l}p_{l}\sum_{i}\tr R^{l}_{\,i}B^{l}_{\,i} \rho_{l} {B^{l}_{\,i}}^{\dagger}
    {R^{l}_{\,i}}^{\dagger} 
      \label{t5}\\
    &\leq
      \sum_{l}p_{l}\sum_{ij}\tr R^{l}_{\,i}B^{l}_{\,j}\rho_{l}{B^{l}_{\,j}}^{\dagger}{R^{l}_{\,i}}^{\dagger}
      \label{t6}\\
    &=
      \sum_{l}p_{l}\tr\mathcal{R}(\mathcal{B}(\rho_{l}))
      \label{t7}\\
    &= \tr \mathcal{R}(\mathcal{B}(\rho))
      \notag\\
    &\leq \tr \mathcal{R}(\mathcal{A}(\rho)) = 1.
      \label{eq13}
  \end{align}
  Line~\eqref{t1} substitutes the definition and applies cyclicity of the trace
  and \([\rho,\rho_{l}]=0\). Line~\eqref{t2} substitutes according to Eq.~\eqref{eqrb}
  with the \(l\)'th operator expansions. Line~\eqref{t3} uses cyclicity of the
  trace to prepare for Line~\eqref{t4}'s application of the operator Schwarz inequality.
  Line~\eqref{t5} follows from \(\tr(\rho_{l})=1\).
  Line~\eqref{t6} expands the sum by adding positive terms. Line~\eqref{t7}
  recognizes the sum of the expression under the trace as the application
  of \(\mathcal{R}\) and \(\mathcal{B}\). The inequality used in the last line 
  was noted earlier and established with Eq.~\ref{eqbarho}.
\end{proof}

Two important special cases of Thm.~\ref{thm2} are in the following corollary.
\begin{corollary}
  Let \(\mathcal{R}\) be a
  trace-preserving completely positive map. Then
  \(F_{\rm cl}(\rho,\mathcal{R}\mathcal{A})
  \leq \sqrt{F_{\rm cl}(\rho,\mathcal{R}_{\mathcal{A},\rho}\mathcal{A})}\)
  and \(F_{e}(\rho,\mathcal{R}\mathcal{A})\leq
  \sqrt{F_{e}(\rho,\mathcal{R}_{\mathcal{A},\rho}\mathcal{A})}\).
\end{corollary}
When the members of the input ensemble \(\rho_{i}\) do not commute, we
do not know whether \(\mathcal{R}_{\mathcal{A},\rho}\) for
\(\rho=\sum_{i}p_{i}\rho_{i}\) is still near-optimal.

\section{Relationship to the ``pretty good measurement''}
\label{sec5}

The above-presented analysis of the fidelity of reversal makes it
clear that \(\mathcal{R}_{\mathcal{A},\rho}\) provides a method for
distinguishing, with close to optimal average error, density matrices
from the ensemble
\(\{p_{j},\hat{\rho}_{j}\}_{j}\), where \(\hat{\rho}_{j}\coloneqq\mathcal{A}(\ketbra{j})\)
and \(\rho=\sum_{j}p_{j}\ketbra{j}\).
This
provides a near-optimal method for distinguishing density matrices in an arbitrary
ensemble, since any ensemble \(\{p_{j},\hat{\rho}_{j}\}_{j}\)
may be constructed by an operation
\begin{align}
  \mathcal{A}\sim\{\sqrt{\lambda_{ij}}\ketbra{v_{ij}}{j}\}_{ij},
  \label{eq15}
\end{align}
where \(\hat{\rho}_{j}=\sum_{i}\lambda_{ij}\ketbra{v_{ij}}\)
are the spectral decompositions of the density
matrices to be distinguished. The operation \(\mathcal{A}\) may be thought of as measuring in the
orthogonal basis \(\ket{j}\), and then
producing the corresponding \(\hat{\rho}_{j}\), for example, by randomly applying with
probabilities \(\lambda_{ij}\) unitary rotations taking \(\ket{j}\) to \(\ket{v_{ij}}\).
With this operation \(\mathcal{A}\),
\begin{align}
  \mathcal{R}_{\mathcal{A},\rho}\sim
  \{R_{ij}\}_{ij} = \qty{\sqrt{p_{j}}\sqrt{\lambda_{ij}}\ketbra{j}{v_{ij}}\rho_{\text{out}}^{-1/2}}_{ij},
  \label{eq16}
\end{align}
where \(\rho_{\text{out}}=\sum_{j}p_{j}\hat{\rho}_{j}\).  The ``pretty
good measurement'' (PGM) was introduced by Holevo~\cite{Holevo1978}
for the case of linearly independent pure states, in which case the
PGM is a measurement of orthogonal projectors, and as Holevo showed,
the optimal such measurement.  The term ``pretty good measurement''
is from Ref.~\cite{Hausladen1996}.  For an ensemble of unnormalized
density matrices \(\rho_{j}\coloneqq p_{j}\hat{\rho}_{j}\) so that
\(\rho_{\text{out}}=\sum_{j}\rho_{j}\) is a normalized density
operator, the PGM is defined by the set of operators consisting of the
\begin{align}
  X_{j}\coloneqq \rho_{\text{out}}^{-1/2}\rho_{j}\rho_{\text{out}}^{-1/2}.
\end{align}
For pure states \(\rho_{j}\) with \(\rho_{j}\propto\ketbra{v_{j}}\),
these are just the operators corresponding to the
``\(\rho\)-distorted''~\cite{Hughston1993} states
\(\rho_{\text{out}}^{-1/2}\ket{v_{j}}\).  Note that for a doubly
indexed ensemble consisting of unnormalized states \(\rho_{ij}\) with
\(\sum_{i}\rho_{ij}=\rho_{j}\), we have \(\sum_{i}X_{ij}=X_{j}\),
where the \(X_{j}\) are the PGM for the ensemble consisting the
\(\rho_{j}\).  The operation in Eq.~\eqref{eq16} may be viewed, via
the given representation, as performing the PGM for the ensemble
consisting of the unnormalized states
\(\sqrt{p_{j}}\sqrt{\lambda_{ij}}\ket{v_{ij}}\), and returning
\(\ket{j}\) when the measurement result \(i,j\) is obtained. Indeed,
for this ensemble \({R_{ij}}^{\dagger}R_{ij}=X_{ij}\), and therefore
\(\sum_{i}{R_{ij}}^\dagger R_{ij}=X_{j}\).  Thus the operation may
also be viewed as doing the PGM for the \(\rho_{j}\), and returning
\(\ket{j}\) when the measurement result is \(j\) .  Since the
\(\ket{j}\) are orthogonal, this is ``classical information'': the
label \(j\) is viewed as the estimate of which \(j\) was input.
However, a given ensemble consisting of the \(\rho_{j}\) may in general
arise from orthogonal states \(\ket{j}\) by actions of channels
different from the ``classicizing'' one of Eq~\eqref{eq15}, which
completely decoheres the orthogonal states \(\ket{j}\) before
producing \(\hat{\rho}_{j}\). For example, if the \(\hat{\rho}_{j}\)
are orthogonal and pure they may be produced either by measurement in
the basis \(\ket{j}\) followed by an appropriate unitary operator
\(U\), or by applying \(U\) without prior measurement. In the first
case quantum coherence is completely destroyed, while in the second
case it is perfectly preserved. When the channel producing the
\(\hat{\rho}_{j}\) is not of the form in Eq.~\eqref{eq15}, the
reversal operation \(\mathcal{R}_{\mathcal{A},\rho}\) will be
different from the one in Eq.~\eqref{eq16}. Although the operation of
Eq.~\eqref{eq16} still gives near-optimal classical fidelity, it will
not necessarily give good entanglement fidelity, since in some sense
it decoheres the states \(\hat{\rho}_{j}\) .
\(\mathcal{R}_{\mathcal{A},\rho}\), however, has near-optimal
entanglement fidelity while retaining near-optimal classical
fidelity. \(\mathcal{R}_{\mathcal{A},\rho}\) thus takes advantage of
whatever coherence remains between the \(\hat{\rho}_{j}\); it avoids
decohering the \(\hat{\rho}_{j}\) if the channel has not decohered
them already.

\section{A bound on the classical fidelity of reversal}
\label{sec6}

To bound a fidelity of reversal it is sufficient to bound the fidelity
for the near optimal reversal operation and apply
Thm.~\ref{thm2}. Here we have a look at such bounds for classical
fidelities of reversal for \(\mathcal{A}\) of the form in
Eq.~\eqref{eq15}. In this case, the classical fidelities are average
probabilities of success for measurements that attempt to infer which
of the eigenprojectors \(\ketbra{j}\) of the input state
\(\rho=\sum_{j}p_{j}\ketbra{j}\) was actually transmitted. The
expression for the PGM gives the following bound on the optimal
probability of success \(F_{\text{cl}}\), with the definitions of Sec.~\ref{sec5}.
\begin{align}
  F_{\text{cl}}^{2}
  &\geq F_{\text{cl}}(\rho,\mathcal{R}_{\mathcal{A},\rho}\mathcal{A})
    \notag\\
  &=\sum_{j}\tr \rho_{\text{out}}^{-1/2}\rho_{j}\rho_{\text{out}}^{-1/2}\rho_{j}
    \notag\\
  &= 1-\sum_{i,j:i\ne j}\tr \rho_{\text{out}}^{-1/2}\rho_{i}\rho_{\text{out}}^{-1/2}\rho_{j},
    \label{eqfclr}
\end{align}
where we used the identity
\(\sum_{i,j}\tr
\rho_{\text{out}}^{-1/2}\rho_{i}\rho_{\text{out}}^{-1/2}\rho_{j}=1\).
Thus the probability of error \(E_{\text{cl}}\) is bounded above by
\(2\sum_{i,j:i\ne j}\tr
\rho_{\text{out}}^{-1/2}\rho_{i}\rho_{\text{out}}^{-1/2}\rho_{j}\),
which is a multiple of the sum of the Hilbert–Schmidt inner products
of the different
\(\rho_{\text{out}}^{-1/4}\rho_{j}\rho_{\text{out}}^{-1/4}\).  When
\(\rho_{\text{out}}\) is proportional to a projection, this sum can be
easy to estimate. An often used measure of overlap between density
matrices is the Bures–Uhlmann
fidelity~\cite{Bures1969,Uhlmann1976}. This measure depends only on
the pair of density matrices, and is defined by
\(F_{\text{BU}}(\sigma_{1},\sigma_{2})\coloneqq
\tr\sqrt{\sigma_{1}^{1/2}\sigma_{2}\sigma_{1}^{1/2}}\). The expression
for the optimal reversal given in Eq.~\eqref{eq16} can be used to
derive a bound on the probability of error in terms of the
Bures–Uhlmann fidelities.

\begin{theorem}\label{thm3}
  \begin{align}
    F_{\rm cl}^{2}\geq
    1-\sum_{i,j:i\ne j}\sqrt{p_{i}p_{j}}F_{\rm BU}(\hat{\rho}_{i},\hat{\rho}_{j}).
  \end{align}
\end{theorem}

\begin{proof}
  Let \(A_{j}\) be the matrix whose \(i\)'th column is
  \(\sqrt{p_{j}}\sqrt{\lambda_{ij}}\ket{v_{ij}}\), and \(A\) the
  matrix with one block row whose \(j\)'th block is \(A_{j}\).  Then
  \(AA^{\dagger}=\sum_{j}A_{j}{A_{j}}^{\dagger} =
  \sum_{j}\rho_{j}=\rho_{\text{out}}\). Let \(R_{j}\) be the matrix
  whose \(i\)'th row is
  \(\sqrt{p_{j}}\sqrt{\lambda_{ij}}\bra{v_{ij}}\rho_{\text{out}}^{-1/2}\),
  and \(R\) the matrix with one block column vector whose \(j\)'th
  block is \(R_{j}\). \(R\) can be viewed as an explicit array form of
  \(\mathcal{R}_{\mathcal{A},\rho}\). Since
  \(R=A^{\dagger}(AA^{\dagger})^{-1/2}\), \(R\) has the property that
  \(RA\) is positive semidefinite.  We remark that this gives an
  alternative approach to defining
  \(\mathcal{R}_{\mathcal{A},\rho}\).

  The matrix \(RA\) has a natural block structure that mirrors that
  used to define \(R\) and \(A\).  The block at block position \(k,l\) in \(RA\) is 
  \((RA)_{kl}=R_{k}A_{l}=A_{k}^{\dagger}\rho_{\text{out}}^{-1/2}A_{l}\).
  We need the following identity:
  \begin{align}
    |RA|_{2}^{2}
    &= \sum_{kl}\tr\, {(RA)_{kl}}^{\dagger}(RA)_{kl}
      \notag\\
    &= \tr \sum_{kl}
      {A_{l}}^{\dagger}\rho_{\text{out}}^{-1/2}A_{k}{A_{k}}^{\dagger}\rho_{\text{out}}^{-1/2}A_{l}
      \notag\\
    &= \tr \sum_{l}{A_{l}}^{\dagger}\rho_{\text{out}}^{-1/2}\sum_{k}\rho_{k}\,
      \rho_{\text{out}}^{-1/2}A_{l}
      \notag\\
    &=\tr \sum_{l}{A_{l}}^{\dagger}\rho_{\text{out}}^{-1/2}\rho_{\text{out}}\rho_{\text{out}}^{-1/2}A_{l}
      \notag\\
    &=
      \tr \sum_{l} {A_{l}}^{\dagger}A_{l}
    = \tr\sum_{l}A_{l}{A_{l}}^{\dagger}= \tr \rho_{\text{out}} = 1.
  \end{align}
  The classical fidelity
  \(F_{\text{cl}}(\rho,\mathcal{R}_{\mathcal{A},\rho}\mathcal{A})\) is the sum of
  the squared Frobenius norms
  \(|(RA)_{jj}|_{2}^{2}=\tr(RA)_{jj}(RA)_{jj}^{\dagger}\) of
  the diagonal blocks of \(RA\), as shown by applying the second line
  of Eq.~\eqref{eqfclr}:
  \begin{align}
    F_{\text{cl}}(\rho,\mathcal{R}_{\mathcal{A},\rho}\mathcal{A})
    &= \sum_{j}\tr \rho_{\text{out}}^{-1/2}\rho_{j}\rho_{\text{out}}^{-1/2}\rho_{j}
      \notag\\
    &= \sum_{j} \tr (AA^{\dagger})^{-1/2}A_{j}A_{j}^{\dagger}
      (AA^{\dagger})^{-1/2}A_{j}A_{j}^{\dagger}
      \notag\\
    &= \sum_{j} \tr A_{j}^{\dagger}(AA^{\dagger})^{-1/2}A_{j}A_{j}^{\dagger}
      (AA^{\dagger})^{-1/2}A_{j}
      \notag\\
    &= \sum_{j}\tr (R_{j}A_{j})(A_{j}^{\dagger} R_{j}^{\dagger})
      \notag\\
    &= \sum_{j}\tr (RA)_{jj}(RA)_{jj}^{\dagger}.
  \end{align}
  Since \(|RA|_{2}^{2}\) is one, it suffices to estimate the sum of
  the squared Frobenius norms of the off-diagonal blocks of \(RA\) to
  bound the optimal \(F_{\text{cl}}\). To do so, let
  \(B=(RA)^{\dagger}(RA) = (RA)^{2}=A^{\dagger}A\). The squared
  Frobenius norm of the off-diagonal block at block position \(k,l\)
  of \(RA\) is the trace of the block \(B_{kl}={A_{k}}^{\dagger}A_{l}\) at block position
  \(k,l\) of \(B\).  Since \(A_{k}{A_{k}}^{\dagger}=\rho_{k}\) and by
  polar decomposition of \(A_{l}\),
  \begin{align}
    {B_{kl}}^{\dagger}B_{kl} = {A_{l}}^{\dagger}\rho_{k}A_{l}= U \rho_{l}^{1/2}\rho_{k}\rho_{l}^{1/2}U^{\dagger}
  \end{align}
  for some unitary operator \(U\) depending on \(l\). Consequently,
  the \(L_{1}\)-norm of \(B_{kl}\), defined by
  \(|B_{kl}|_{1}= \tr \sqrt{{B_{kl}}^{\dagger}B_{kl}}\) is
  \(\sqrt{p_{k}p_{l}}F_{\text{BU}}(\hat{\rho}_{k},\hat{\rho}_{l})\).
  It therefore suffices to relate the Frobenius norms of the
  off-diagonal blocks of a positive semidefinite matrix to the
  \(L_{1}\) norms of the off-diagonal blocks of its square, via the
  following lemma.

  \begin{lemma}\label{lem4}
    Let
    \begin{align}
      M=\begin{pmatrix}a & b^{\dagger}\\b & c
      \end{pmatrix}
    \end{align}
    be positive semidefinite, with \(a,b,c\) matrices. Write
    \begin{align}
      M^{1/2}=\begin{pmatrix}x&y^{\dagger}\\y&z
      \end{pmatrix}
    \end{align}
    with the same block structure. Then \(|y|_{2}^{2}\leq |b|_{1}\).
  \end{lemma}

  \begin{proof}
    Without loss of generality, assume that \(y\) is non-negative diagonal.
    Otherwise, with a block-diagonal unitary
    \begin{align}
      U=\begin{pmatrix}u&0\\0&v
      \end{pmatrix}
    \end{align}
    with \(u\) and \(v\) chosen to implement the singular value
    decomposition of \(y\) , we may transform \(M\) and \(M^{1/2}\) so
    that \(y\) is a diagonal matrix with non-negative diagonal
    entries.  For rectangular \(y\) , the upper or left-hand square
    portion is diagonalized. This does not affect the norms, since
    \(U\) transforms blocks independently and the \(L_{1}\) and
    Frobenius norms are both unitarily invariant. Let \(y_{i}\) be the
    diagonal entries of \(y\). Note that \(b= yx+zy\) and
    \(|b|_{1}\geq \tr b\), see Ref.~\cite{HornJohnson1985},
    p. 432. Now \(\tr(yx+zy)=\sum_{i}y_{i}(x_{ii}+z_{ii})\).  By the
    positivity of \(M^{1/2}\), \(y_{i}^{2}\leq x_{ii}z_{ii}\), so
    \(y_{i}\) is less than at least one of \(x_{ii},z_{ii}\).  Thus
    \(|y|_{2}^{2}=\sum_{i}y_{i}^{2}\leq \tr b \leq |b|_{1}\), as
    desired.  
  \end{proof}
  \noindent Remark: Near the end of the above proof, because
  \(2y_{i} = 2\sqrt{x_{ii}}\sqrt{z_{ii}}\leq x_{ii}+z_{ii}\), the
  inequality \(y_{i}^{2}\leq y_{i}(x_{ii}+z_{ii})/2\) could have been
  used to obtain a stronger bound.
    
  To complete the proof of Thm.~\ref{thm3}, consider first the
  \(2\times 2\) block decomposition of \(RA\) and \(B\) with upper
  left-hand block \((RA)_{11}\) and \(B_{11}\). By Lem.~\ref{lem4},
  the squared Frobenius norm of the first block row and column
  excluding \((RA)_{11}\) is upper bounded by the sum of the \(L_{1}\) norms
  of the corresponding block row and column in \(B\). By subadditivity
  of the norm, this is at most
  \(\sum_{i>1}\qty(|B_{1i}|_{1}+|B_{i1}|_{1})\).  After a suitable
  permutation, the same argument applies to the row and column
  determined by \(B_{ii}\), for each \(i\). The proof of the theorem
  then follows by summing over the resulting inequalities and noting
  that each off-diagonal block occurs twice on both sides.
\end{proof}

\section{Examples and applications}
\label{sec7}

By a slight extension of an example already given,
\(\mathcal{R}_{\mathcal{A},\rho}\) is optimal when \(\mathcal{A}\) is
perfectly reversible on some code subspace and \(\rho\) is a state
with no support outside the code.  Applications to reversing other
simple quantum operations may be instructive. For instance, consider a
qubit depolarizing channel \(\mathcal{A}\) whose operator decomposition
consists of \((1-p)^{1/2}I\) and the \((p/3)^{1/2}\sigma_{i}\) for
\(i=1,2,3\) with \(\sigma_{i}\) being the Pauli operators.
For \(\rho=I/2\),  \(\mathcal{R}_{\mathcal{A},\rho}\) is the same
depolarizing channel, whereas the optimal reversal is to do
nothing. This case saturates the inequality of Thm.~\ref{thm3}.

Due to its near optimality, the reversal operation
\(\mathcal{R}_{\mathcal{A},\rho}\) can be used in any situation where
classical or quantum information has been corrupted by noise with
known behavior.  \(\mathcal{R}_{\mathcal{A},\rho}\) has a simple
definition, but whether it or a good approximation can be implemented
efficiently depends on the details of the situation.  Because its
error is at most twice the optimum, it can be used as a theoretical
tool to obtain upper bounds on the achievable fidelities in a given
situation regardless of whether or not it can be efficiently
implemented.  The upper bounds can then be compared to the fidelity
achieved by simpler algorithms. An example of this occurs in the use
of stabilizer codes for quantum error-correction. When the noise model
is independent and depolarizing, classical coding theory immediately
suggests a combinatorially straightforward error-correction algorithm
based on maximum likelihood error syndrome decoding. Comparing this
method to \(\mathcal{R}_{\mathcal{A},\rho}\) suggests itself as a
fruitful path of investigation with applications to asymptotic bounds
in quantum coding theory~\cite{Ashikhmin2000a,Ashikhmin2000b}. More
generally, for any encoding scheme capable of transmitting quantum
information through a given channel at a given rate when appropriate
decoding is used, \(\mathcal{R}_{\mathcal{A},\rho}\) may be used to
provide such a decoding, with \(\mathcal{A}\) taken to be the
concatenation of encoding and noise.

Another application is to query complexity for quantum oracles. In this case,
we are given a quantum black box implementing an unknown quantum
operation from some set. A simple method for attempting to determine
which operation we are given is to apply it to copies of some input
state and attempt to distinguish the output state. A bound on the
probability of success can then be obtained by using bounds such as
the one of Thm.~\ref{thm3}.  This was how the fact that the hidden
subgroup problem has low query complexity was first
realized~\cite{Ettinger1999}.

\acknowledgments{ The authors thank the following for support: The ONR
  (N00014-93-1-0116, H.B.), the NSF (PHY-9722614, H.B.), the ISI
  Foundation (Turin, Italy, H.B.), Elsag-Bailey (H.B.), the ITP at UC
  Santa Barbara (NSF PHY94-07194, H.B. and E.K.), the NSA (E.K.) and
  the DOE (W-7405-ENG-36, E.K.). We thank Mohammad Alhejji for
  pointing out that the original Eq.~\eqref{eq13} did not account for
  the probabilities \(p_{l}\), and when these probabilities are
  included, the inequalities fail to hold.  As a result, in the
  originally published version, Ref.~\cite{barnum}, Thm.~\ref{thm2}
  was established for entanglement fidelity but not for average
  entanglement fidelity.  }

\end{document}